\documentclass[letterpaper, 10 pt, conference]{ieeeconf}  

\IEEEoverridecommandlockouts                              %
\usepackage{graphics} 
\usepackage{epsfig} 
\usepackage{mathptmx} 
\usepackage{times} 
\usepackage{amsmath} 
\usepackage{amssymb}  
\usepackage{balance}

\newtheorem{remark}{Remark}
\title{\LARGE \bf
Multi-layer barrier adaptation of the discrete-time super-twisting controller
}

\author{Antoine Thibault Vi{\'e}, Leonid Fridman, Roberto Galeazzi and Dimitrios Papageorgiou
	\thanks{A. T. Vi{\'e}, R. Galeazzi and D. Papageorgiou are with the Department of Electrical and Photonics Engineering, Technical University of Denmark, Elektrovej 326, Kgs Lyngby, 2800, Denmark
		{\tt\small anthv,roga,dimpa@dtu.dk}}%
	\thanks{L. Fridman is with Facultad de Ingenieria, Universidad Nacional Autonoma de Mexico,  Mexico City, 04510, Mexico
		{\tt\small lfridman@unam.mx}}%
}

\begin{document}

\maketitle
\thispagestyle{empty}
\pagestyle{empty}

\begin{abstract}
In digital sliding mode control implementations, discretization-induced chattering and inter-sample ``blindness" can severely degrade the closed-loop performance, especially in case of fast perturbations. This paper addresses these challenges for a discrete-time implementation of the super-twisting sliding mode controller. Building upon recent results on barrier-function-modulated super-twisting algorithms, a nested architecture employing multiple barriers is discretized using an eigenvalue-based exact matching approach. The resulting discrete-time controller preserves the adaptive and robustness properties established in continuous time, while ensuring consistent stability behavior at the sampling level. The proposed framework is validated through numerical simulations. 
The results highlight the effectiveness of multi-layer barrier adaptation for discrete-time sliding mode control applications.

\end{abstract}

\section{Introduction}
Sliding mode control (SMC) has long been recognized as a powerful methodology for the robust control of systems subject to matched uncertainties and external disturbances \cite{slotine1991applied,utkin1992scope}. Among several higher-order sliding mode strategies, the super-twisting algorithm (STA) has attracted particular attention due to its ability to enforce finite-time convergence of the sliding variable without requiring direct measurement of its derivative \cite{moreno_lyapunov_2012}, as well as for featuring reduced chattering compared to first-order sliding modes. These properties have rendered the STA especially attractive for practical applications involving noisy measurements and uncertain dynamics.

Recent developments have focused on enhancing the adaptability of the super-twisting controller by introducing state-dependent gain modulation mechanisms \cite{shtessel_super-twisting_2010,shtessel_novel_2012,moreno_adaptive_2015,utkin_adaptive_2013}. In particular, the incorporation of barrier functions into SMC architectures has been shown to guarantee Uniform Ultimate Boundedness (UUB) of the closed-loop trajectories \cite{gonzalez_final_2024} while enabling performance recovery in the presence of unknown but bounded perturbations \cite{obeid_barrier_2020}. Such barrier-based formulations exploit the sliding variable to dynamically adjust control gains, thereby avoiding excessive conservatism.

However, in digital implementations, the switching frequency is inherently constrained by the nonzero sampling period. As a consequence, the so-called discretization-induced chattering phenomenon arises, characterized by self-sustained oscillations in the system output and state variables, which ultimately degrade closed-loop control performance. \cite{koch_discrete_2016,koch_discrete-time_2019,brogliato_implicit_2018,salgado_control_2016}. Moreover, when implemented in discrete time, super-twisting controllers suffer from an inherent inter-sample blindness. This may significantly degrade closed-loop performance and compromise robustness if not properly addressed. It has recently been shown that nested architectures based on multiple barrier functions can alleviate this issue by introducing layered gain adaptation mechanisms that respond differently across regions of the state space \cite{thibault2025multi}.

Motivated by these observations, the present work focuses on the discrete-time realization of a multi-layer barrier function–based adaptive super-twisting controller. An eigenvalue-based discretization technique \cite{koch_discrete-time_2019} is employed to construct a discrete-time equivalent of the continuous-time closed-loop dynamics while preserving stability properties. Unlike conventional discretization approaches, the method explicitly matches the continuous-time eigenstructure, thereby ensuring consistency between continuous and discrete implementations.


The effectiveness of the proposed control strategy is demonstrated through simulations. The results confirm that the multi-layer barrier adaptation significantly improves tracking accuracy and robustness under uncertainty and inter-sample blindedness.

The contributions of the present work can be summarized as follows:
\begin{itemize}
    \item Discrete-time formulation of multi-layer barrier STA
    \item Eigenvalue-based discretization preserving adaptive behavior.
    \item Sensitivity analysis of the proposed scheme according to its key tuning parameters and the sampling time
\end{itemize}

The remainder of the paper is structured as follows: Section \ref{sec:problemformulation} presents the problem at hand along with notation conventions. Section \ref{sec:MultiLayer} details the discrete-time multi-layered algorithm. A simulation-based sensitivity analysis with respect to the parameters of the algorithm is provided in Section \ref{sec:Simulations}. 
Finally, conclusions are drawn in Section \ref{sec:Conclusions}, where future works is also discussed.

\section{Preliminaries} \label{sec:problemformulation}
The system under consideration is described by a perturbed integrator of the form
\begin{equation}
\label{eq:system}
\dot{s} = u + d(t),
\end{equation}
where $s,u \in \mathbb{R}$ and $d : \mathbb{R}^+ \rightarrow \mathbb{R}$.. Any system of relative degree $\varrho$ can be described by \eqref{eq:system} based on an appropriate selection of a sliding-variable (corresponding to an $(\varrho - 1)$-dimensional sliding manifold) and equivalent control \cite{slotine1991applied}.

Within the non-homogeneous super-twisting sliding mode control (STSMC) framework, and using the compact notation $\lfloor s \rceil^a \triangleq |s|^a \mathrm{sgn}(s)$, the control law is defined as
\begin{subequations} \label{eq:HomogeneousSTSMC}
\begin{align}
u &= -k_1 \lfloor s \rceil^{\alpha} + v, \\
\dot{v} &= -k_2 \lfloor s \rceil^{0},
\end{align}
\end{subequations}
where $k_1,k_2 \in \mathbb{R}^{+}$and $\alpha\in \mathbb{R}^{*+}$.

The resulting closed-loop dynamics, interpreted in the Filippov sense, are expressed as
\begin{subequations}
\label{eq:STA}
\begin{align}
\dot{s} &= -k_1 \lfloor s \rceil^\alpha + \phi, \\
\dot{\phi} &= -k_2 \lfloor s \rceil^0 + \delta(t),
\end{align}
\end{subequations}
where $\delta(t) \triangleq \dot{d}(t)$. The perturbation $\delta(t)$ is assumed to be bounded, such that there exists a constant $\Delta > 0$ satisfying $\lvert \delta(t) \rvert \leq \Delta$ for all $t \in \mathbb{R}^+$, while $\Delta$ is not known a priori.

It has been shown in \cite{moreno_strict_2014} that the incorporation of barrier functions into sliding mode control architectures ensures UUB of the closed-loop trajectories. 
Moreover, it has been shown \cite{thibault2025multi} that a semi-definite barrier function of the form: 
\begin{subequations}
    \begin{align}
        k_{1,\alpha}(s) &= \frac{\lvert s \rvert}{(\epsilon - \lvert s \rvert)^{\alpha+1}}, \\
        k_{2,\alpha}(s) &= k_{1,\alpha}(s)^2,
    \end{align}
    \label{eq:BF}
\end{subequations}
guarantees UUB of the closed-loop solutions, where $ k_{1,\alpha}(s),  k_{2,\alpha}(s)$ are the gains of the system \eqref{eq:STA}.

For digital implementation, the continuous-time dynamics \eqref{eq:STA} are discretized using a constant sampling period $T_s \in \mathbb{R}^+$, applying a zero-order hold (ZOH) both on the control input and the sliding variable, and finally, by considering a piecewise-constant approximation of $\delta(t)$ over each sampling interval. Hence, a first-order forward Euler discretization of \eqref{eq:STA} yields the discrete-time system
\begin{subequations}
    \begin{align}
        s_{k+1} &= s_k + T_s(-k_{1,\alpha}(s_k)\lfloor s_k \rceil^\alpha + \phi_k), \\
        \phi_{k+1} &= \phi_k -T_s k_{2,\alpha}(s_k) \lfloor s_k \rceil^0 + T_s\delta(kT_s),
    \end{align}
\end{subequations}
which coincides with the discrete-time super-twisting approximation proposed in \cite{koch_discrete-time_2019}.

\section{Multi-layered barrier function adaptation in discrete time} \label{sec:MultiLayer}
It has been demonstrated in \cite{thibault2025multi} that, due to the fact that the controller operates without access to intra-sample information, a nested architecture incorporating multiple barrier functions yields improved closed-loop performance. In this context, the objective of the present section is to reuse the adaptive gain modulation strategy proposed in \cite{thibault2025multi}, which is recalled below for completeness: 
\begin{align} \label{eq:A0AiModulation}
	&A_0 : \begin{cases}
		\dot{k}_{1,\alpha,d} = \displaystyle {k_{1,\alpha,d}}/{\lvert \dot{s} \rvert} \\
		\dot{k}_{2,\alpha,d} = \displaystyle \frac{k_{2,\alpha,d}}{2 \lvert s \rvert^{1-\alpha}} \\
		k_{1,\alpha} = k_{1,\alpha,d} \\
		k_{2,\alpha} = k_{2,\alpha,d}
	\end{cases}, A_i : \begin{cases}
		\dot{k}_{1,\alpha,d} = -k_{1,\alpha,d} \\
		\dot{k}_{2,\alpha,d} = -k_{2,\alpha,d} \\
		k_{1,\alpha} = \displaystyle \frac{\lvert s \rvert}{\left(\epsilon_i - \lvert s \rvert\right)^{\alpha+1}} \\
		k_{2,\alpha} = k_{1,\alpha}^2
	\end{cases},
\end{align} 
A one-step memory variable $a$ is introduced to retain the previously selected scheme. The resulting nested switching algorithm can be expressed as follows:
\begin{align*} 
	\begin{cases}
		A_0 \text{ if } \lvert s \rvert > \epsilon_N \vee \left( \epsilon_N > \lvert s \rvert > \epsilon_1 \wedge  a = A_0 \right) \\
		A_i \text{ if } \epsilon_{i-1} < \lvert s \rvert < \epsilon_i  \wedge a \neq A_0\\
		A_1 \text{ if } \lvert s \rvert < \epsilon_1
	\end{cases}
\end{align*}
where $N$ is the number of barriers and the barriers are ordered as $0 < \epsilon_1 < \cdots < \epsilon_N$. The innermost barrier $\epsilon_1$ is associated with the desired target precision, whereas the outermost barrier $\epsilon_N$ defines the limit of acceptable performance. In the general case, the intermediate widths $\epsilon_i$ are uniformly spaced between the innermost and outermost layers. A nested switching logic is employed to interconnect the barrier layers. Specifically, the region $A_i$ corresponding to the current value of $s$, defined by $\lvert s \rvert \in [\epsilon_{i-1},\epsilon_i)$, is identified, and the associated barrier gain is subsequently applied.
\begin{remark}
The stability of the closed-loop system under the nested switching logic described in \eqref{eq:A0AiModulation} is addressed as follows. It has been established in \cite{thibault2025multi} that, for each active barrier layer $A_i$, the barrier-modulated gains $k_{1,\alpha}(s)$ and $k_{2,\alpha}(s)$
defined in \eqref{eq:BF} guarantee UUB of the closed-loop trajectories with respect to the region $\lvert s \rvert\leq \varepsilon_i$. The switching between layers is therefore interpreted as a transition between a family of UUB systems, each associated with a distinct ultimate bound. When, due to inter-sample blindness, the sliding variable $s$ is displaced beyond the innermost boundary $\epsilon_1$ and a higher barrier layer $A_i$ ($i > 1$) is activated, UUB is preserved with respect to the corresponding outer bound $\epsilon_i$. Conversely, when effective disturbance rejection causes the sliding variable to re-enter an inner region, the switching logic transitions to a layer associated with a tighter ultimate bound, thereby recovering finer regulation. Since each subsystem is individually UUB and the switching logic is monotone in the sense that it assigns the tightest admissible barrier consistent with the current state, no destabilizing interaction between layers is introduced by the switching itself. The overall closed-loop system is therefore UUB under arbitrary switching within the nested barrier architecture, with the effective ultimate bound determined by the outermost barrier layer activated during operation.
\end{remark}
\begin{figure}[t] 
	\centering
	\includegraphics[width=0.45\textwidth]{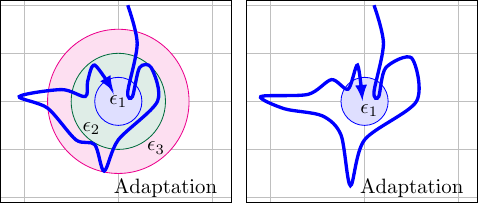}
	\caption{Conceptual illustration of three-layer barrier architecture (left) with step disturbances compared to single-layer barrier function (right). In the latter case the system solely relies on dynamic adaptation when $\vert s \vert > \epsilon_1$.}
	\label{fig:SmultipleLayers}
\end{figure}

The above algorithm is implemented in discrete time using an eigenvalue-based discretization strategy.
The primary motivation for using this scheme is to reduce chattering compared to classical Euler or Tustin discretization methods \cite{koch_discrete-time_2019}.
Since stability in continuous time has already been established in \cite{thibault2025multi}, the analysis is restricted to the discretization procedure described in \cite{koch_discrete-time_2019}. To this end, an alternative representation of the continuous-time dynamics in Eq. \eqref{eq:STA} is adopted, following \cite{reichhartinger_arbitrary-order_2018}. By exploiting the identity
\begin{equation}
    \mathrm{sgn}(s) = \frac{s}{\lvert s \rvert}, \quad \forall s \neq 0
\end{equation}
the system \eqref{eq:STA} can be reformulated in Rosenbrock form as
\begin{equation}
    \dot{x} = M_{\alpha}(s) x + \begin{bmatrix}
        0 \\ \delta(t)
    \end{bmatrix}, \quad \text{if } s \neq 0 
\end{equation}
with $x = \left[ s , \phi\right]^T$ and $M_\alpha: \mathbb{R}^{*} \rightarrow \mathbb{R}^{2\times2}$ is defined as: 
\begin{equation}
    M_\alpha(s) = \begin{bmatrix}
        -k_{1,\alpha} \lvert s\rvert^{\alpha-1} & 1 \\
        -k_{2,\alpha} \lvert s \rvert^{-1} & 0
    \end{bmatrix}.
\end{equation}
The eigenvalues of the matrix $M_\alpha(s)$ are given by
\begin{equation}
    \lambda_{1,2} = \begin{cases}
        - \frac{k_{1,\alpha}\lvert s \rvert^{\alpha-1}}{2} \pm \frac{\sqrt{D}}{2}, \quad \text{if } D \geq 0,\\
         - \frac{k_{1,\alpha}\lvert s \rvert^{\alpha-1}}{2}  \pm
         i\frac{ \sqrt{\lvert D \rvert}}{2}, \quad \text{if } D < 0,
    \end{cases}
\end{equation}
where 
\begin{equation}
    D = k_{1,\alpha}\lvert s \rvert^{2\alpha-2} - 4k_{2,\alpha}\lvert s \rvert^{-1}.
\end{equation}
When the barrier function defined in \eqref{eq:BF} is employed, the eigenvalues can be expressed as
\begin{equation}
     \lambda_{1,2} = \begin{cases}
     -\frac{\lvert s \rvert^\alpha}{2(\epsilon_i - \lvert s \rvert)^{\alpha+1}} \pm \frac{ \lvert s \rvert^\frac{1}{2}\sqrt{\lvert s\rvert^{2\alpha-1} - 4}}{2(\epsilon_i - \lvert s \rvert)^{\alpha+1}}, &\lvert s \rvert^{2\alpha-1} \geq 4 \\
       -\frac{\lvert s \rvert^\alpha}{2(\epsilon_i - \lvert s \rvert)^{\alpha+1}} \pm i\frac{ \lvert s \rvert^\frac{1}{2}\sqrt{\lvert \lvert s\rvert^{2\alpha-1} - 4 \rvert}}{2(\epsilon_i - \lvert s \rvert)^{\alpha+1}},&\text{otherwise}.
    \end{cases}
\end{equation}

To avoid the occurrence of real eigenvalues with positive real parts, the design parameter $\epsilon$ is selected such that $\epsilon < 4^{\frac{1}{2\alpha-1}}$ for $\alpha \neq \tfrac{1}{2}$, while any $\epsilon \in \mathbb{R}^+$ is admissible when $\alpha = \tfrac{1}{2}$. Under this condition, the following asymptotic properties hold for each barrier:
\begin{subequations}
    \begin{align}
        \lim_{\lvert s \rvert \rightarrow 0^+} \lvert \lambda_{1,2} \rvert &= 0 \text{ and } \lim_{\lvert s \rvert \rightarrow \epsilon^-} \lvert \lambda_{1,2} \rvert = \infty,
    \end{align}
\end{subequations} 
The matching discretization scheme is adopted to preserve the structural properties of the continuous-time system.
The discrete-time eigenvalues obtained via exact matching are then given by
\begin{equation}
    \lambda_{1,2}^q = e^{\lambda_{1,2}T_s}.
\end{equation}
The resulting discrete-time system is expressed based on the results in \cite{koch_discrete-time_2019}:
\begin{equation}
    x_{k+1} = M^q_\alpha(s_k) x_k + T_s \begin{bmatrix}
        0 \\ \delta(kT_s)
    \end{bmatrix},
\end{equation}
with
\begin{equation}
    M^q_\alpha(s_k) = \begin{bmatrix}
        \tilde{u}_{1,k} & T_s \\ \tilde{u}_{2,k} & 1
    \end{bmatrix}.
\end{equation}
The coefficients $\tilde{u}_{1,k}$ and $\tilde{u}_{2,k}$ are selected such that $\lambda(M^q_\alpha(s_k)) = \lambda_{1,2}^q$, yielding 
\begin{subequations}
    \begin{align}
        \tilde{u}_{1,k} &= \lambda_1^q + \lambda_2^q - 1, \\
        \tilde{u}_{2,k} &= \frac{1}{T_s} \left( \tilde{u}_{1,k} - \lambda_1^q\lambda_2^q\right).
    \end{align}
\end{subequations}
Finally, the discrete-time implementation of the control law \eqref{eq:HomogeneousSTSMC} is obtained as \cite{koch_discrete-time_2019}
\begin{subequations}
    \begin{align}
        u_k &= \frac{1}{T_s}\left( -s_k + \tilde{u}_{1,k}s_k\right) + v_k ,\\
        v_{k+1} &= v_{k} + \tilde{u}_{2,k}s_k.
    \end{align}
\end{subequations}

\section{Simulations and Sensitivity analysis} \label{sec:Simulations}
Numerical simulations were conducted to evaluate the behavior of the proposed discrete-time multi-layer barrier adaptive super-twisting controller and to assess its sensitivity with respect to the parameters $\alpha$ and the innermost barrier $\epsilon^{-}$. The simulations were designed to highlight the controller’s robustness and its ability to regulate tracking performance under fast and high amplitude perturbations.

The simulated system is a perturbed integrator
\begin{align}
\dot{x} &= u_0 + d
\end{align}
where $x$ must track a reference signal $x_{ref} = 0.1 \sin(5  t)$. Calculating the dynamics of $s \triangleq x - x_{ref}$ with $u_0 = u + \dot{x}_{ref}$ brings the system in the form of Eq. \eqref{eq:system}, where $u$ was generated by the proposed discrete-time super-twisting controller with $\delta = \dot{d}$. The disturbance $d(t) = 10^{3} \sin(15  t)$ was selected to be of both high-amplitude and high-frequency in order to stress the robustness and adaptation mechanisms of the controller.
The controller parameters were selected according to Table~I
\begin{table}[bp]
	\begin{center}
		\begin{tabular}{c|c}
			Parameter & Value \\
			\hline \hline 
			$\alpha$ & $\frac{1}{2}$ \\
			$\epsilon^{-}$ & $1e^{-4}$ \\
			$\epsilon^{+}$ & $\min \left\{1e^{-1}, 1e^{3} \epsilon^-\right\}$ \\
			$T_s$ & $1e^{-5}$ \\
			\hline \hline
		\end{tabular}
	\end{center}
	\caption{Default tunable parameters value}
\end{table}
where $\epsilon^-$ and $\epsilon^+$ are defined as the inner and outer barriers, respectively.

\begin{figure}[h!]
    \centering
    \includegraphics[width=\linewidth]{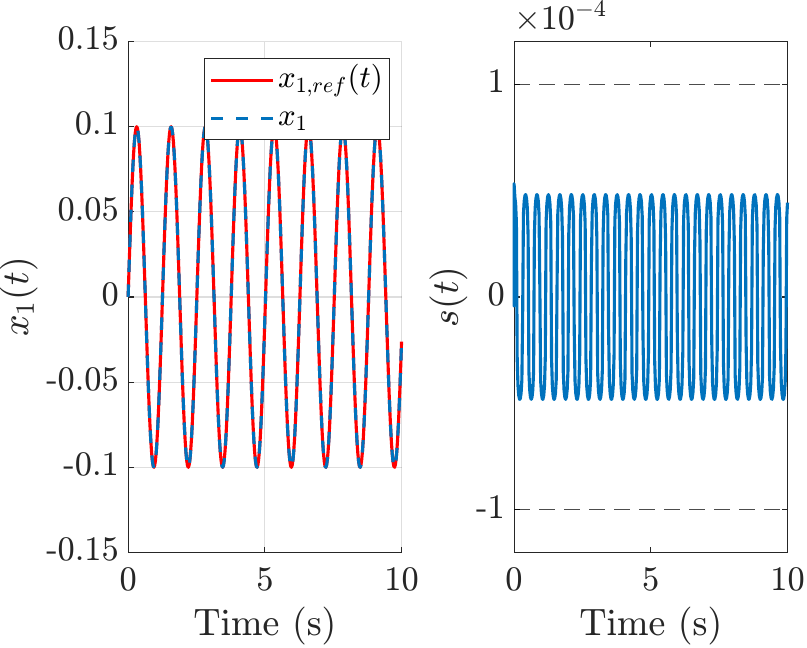}
    \caption{Nominal performance under perturbation $\delta$.}
    \label{fig:NominalDiscrete}
\end{figure}

Fig.~\ref{fig:NominalDiscrete} shows the closed-loop system response. Despite the presence of a large-amplitude disturbance, accurate tracking of the reference signal is achieved. The sliding variable remains confined within the prescribed innermost barrier region showcasing the effectiveness of the discretization scheme.
\subsection{Sensitivity with respect to $\alpha$}
\begin{figure}[t]
    \centering
    \includegraphics[width=\linewidth]{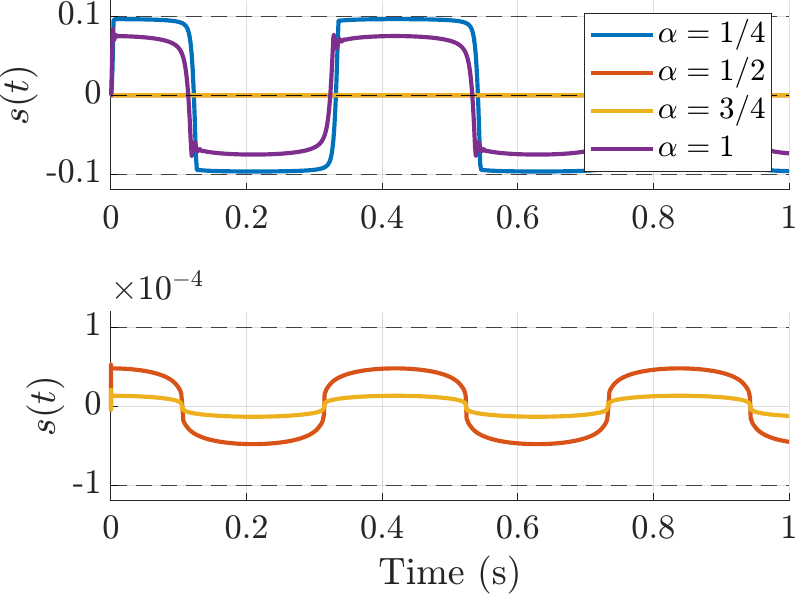}
    \caption{System performance under perturbation $\delta$ for $\alpha=\dfrac{1}{4},\dfrac{1}{2},\dfrac{3}{4},1$.}
    \label{fig:AlphaDiscrete}
\end{figure}

The influence of the parameter $\alpha$ is illustrated in Fig.~\ref{fig:AlphaDiscrete}, where values ranging from $\alpha = 0.25$ to $\alpha = 1$ were considered. For low values of $\alpha$, the closed-loop system becomes under-constrained, resulting in frequent activation of the outermost barrier layer. This behavior leads to reduced steady-state accuracy and increased gain adaptation activity.
Conversely, excessively large values of $\alpha$ impose overly restrictive dynamics, causing the sliding variable to exit the innermost barrier region and rely on higher barrier layers. Prior to this over-constrained regime, however, moderate increases in $\alpha$ relative to the nominal value are observed to improve tracking performance by accelerating convergence toward the sliding manifold without inducing excessive switching. These results demonstrate that $\alpha$ plays a critical role in balancing convergence speed and barrier utilization.

\subsection{Sensitivity with Respect to the innermost barrier $\epsilon^{-}$}
\begin{figure}[t]
    \centering
    \includegraphics[width=\linewidth]{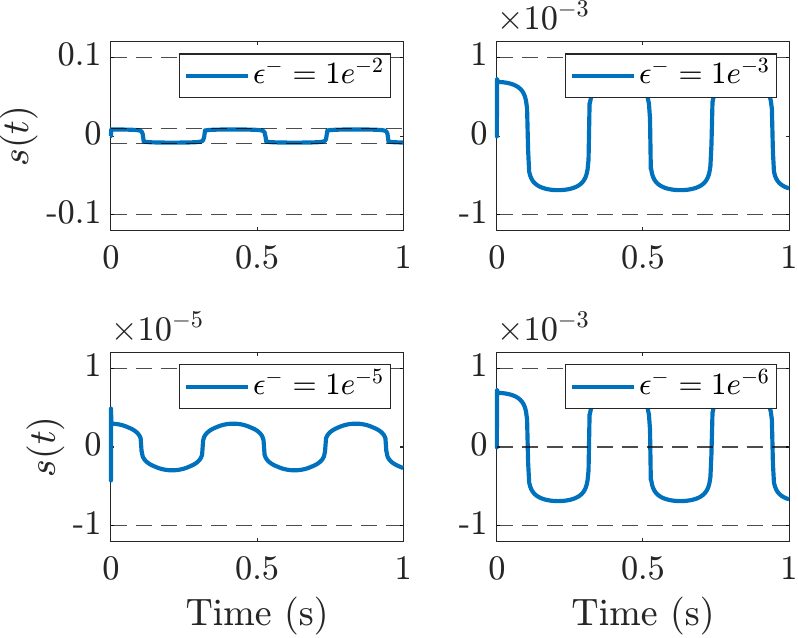}
    \caption{Performance under perturbation $\delta$ for $\epsilon^-=1e^{-2},1e^{-3},1e^{-5}$, $1e^{-6}$.}
    \label{fig:EpsilonDiscrete}
\end{figure}

The sensitivity of the closed-loop response with respect to the inner barrier parameter $\epsilon^{-}$ is shown in Fig.~\ref{fig:EpsilonDiscrete} for four distsinct values:  $1e^{-2},1e^{-3},1e^{-5}$ and $1e^{-6}$. Decreasing $\epsilon^{-}$ leads to a tighter ultimate bound on the sliding variable, thereby enhancing steady-state tracking accuracy. However, excessively small values of $\epsilon^{-}$ result in increased gain magnitudes and more aggressive control action, which may amplify discretization-induced chattering effects or prevent sustained confinement within the innermost barrier region, thus triggering reliance on outer barrier layers.
In contrast, larger values of $\epsilon^{-}$ yield smoother control signals and reduced gain activity, at the expense of relaxed tracking precision. These observations confirm that $\epsilon^{-}$ acts as a key tuning parameter governing the trade-off between steady-state accuracy and control smoothness.

\subsection{Sensitivity with Respect to the Sampling Time $T_s$}

\begin{figure}
    \centering
    \includegraphics[width=\linewidth]{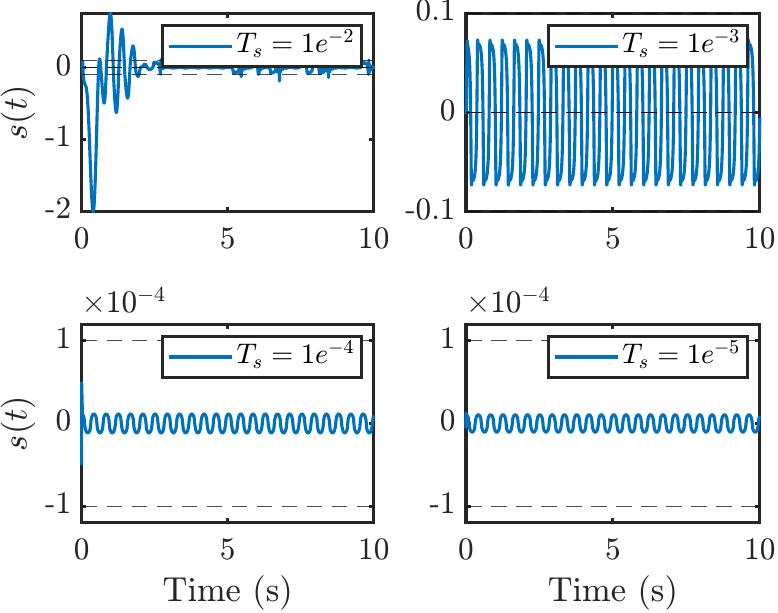}
    \caption{Performance under reduced amplitude perturbation for $T_s = 1e^{-2},1e^{-3},1e^{-4}$ and $1e^{-5}$.}
    \label{fig:S_Ts_scenario}
\end{figure}

In order to assess the effectiveness of the proposed algorithm with respect to the sampling period, the disturbance amplitude is reduced from $1e^3$ to $1e^1$, so that the influence of larger sampling times can be more clearly observed. The results are presented in Fig.~\ref{fig:S_Ts_scenario}. For $T_s =  1e^{-2}$, it is observed that the adaptive algorithm $A_0$ is activated at the onset of the simulation, after which the sliding variable is brought back within the innermost barrier region before being captured by the outermost layer. For $T_s = 1e^{-3}$, the system is found to evolve exclusively within the outermost barrier, as the sliding variable is displaced beyond the innermost boundary during blind inter-sample intervals by the applied perturbation. In contrast, for $T_s =1e^{-4}$ and $T_s = {1e^{-5}}$, the sampling period is sufficiently small relative to the perturbation dynamics such that no inter-sample blind intervals of significance (with respect to the desired accuracy) are encountered, and confinement within the innermost barrier is consistently maintained.

\subsection{Sensitivity with Respect to the Number of Barriers}
\begin{figure}
    \centering
    \includegraphics[width=\linewidth]{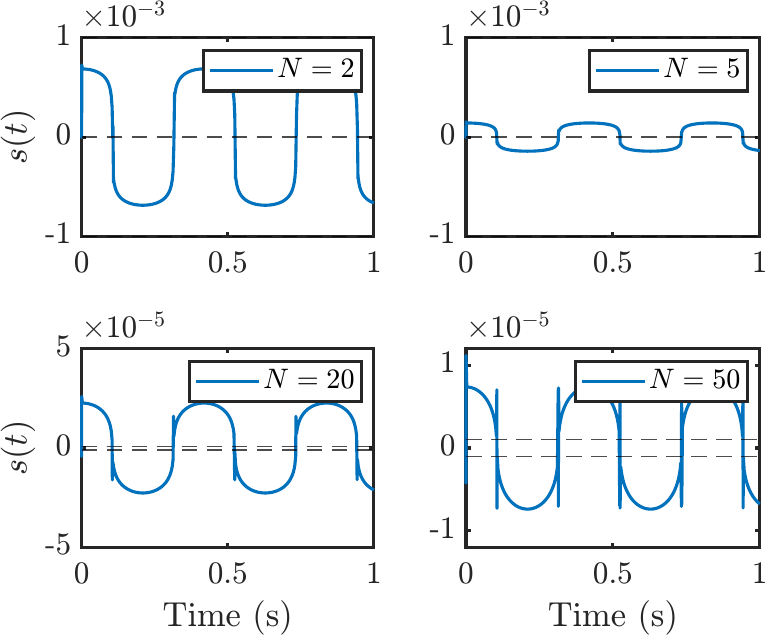}
    \caption{Performance under perturbation for $N=2,5,20, 50$ with $\epsilon^-=1e^{-6}$.}
    \label{fig:S_N_scenario}
\end{figure}

As illustrated in Fig.~\ref{fig:S_N_scenario}, closed-loop performance is observed to improve monotonically with the number of barrier layers. For a fixed sampling period, deviations of the sliding variable are intercepted at earlier barrier layers as $N$ is increased, thereby preventing the trajectory from reaching the outermost boundary. Consequently, perturbation-induced inter-sample deviations are attenuated more rapidly as the number of barriers is increased, resulting in tighter confinement of the sliding variable and improved steady-state accuracy.

\subsection{Cross-Sensitivity: Sampling Time and Number of Barriers}

\begin{figure}
\centering
\includegraphics[width=\linewidth]{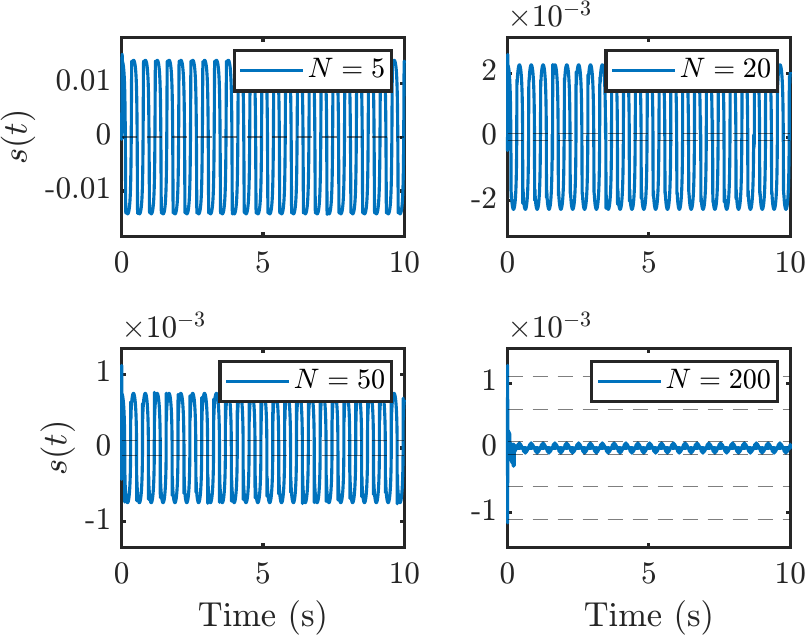}
\caption{Performance under reduced amplitude perturbation for $T_s = 1e^{-3}$, $N = 5, 20, 50, 200$.}
\label{fig:S_N_Ts_scenario}
\end{figure}

The inter-sample blindness inherent to the discretized implementation of the algorithm, as previously discussed and illustrated in Fig.~\ref{fig:S_Ts_scenario}, leads to performance degradation when larger sampling times are employed. This limitation can be mitigated by appropriately adjusting the number of barriers, as demonstrated in Fig.~\ref{fig:S_N_Ts_scenario}. Specifically, it is observed that increasing the number of barriers enhances the likelihood that the sampled state lies within an inner barrier region, thereby improving the overall system performance.

However, it is also observed that a higher number of barriers induces increased chattering, manifested as more frequent switching between barrier layers. This trade-off between performance improvement and chattering must therefore be carefully considered in the design of the control scheme.

\begin{figure}
\centering
\includegraphics[width=\linewidth]{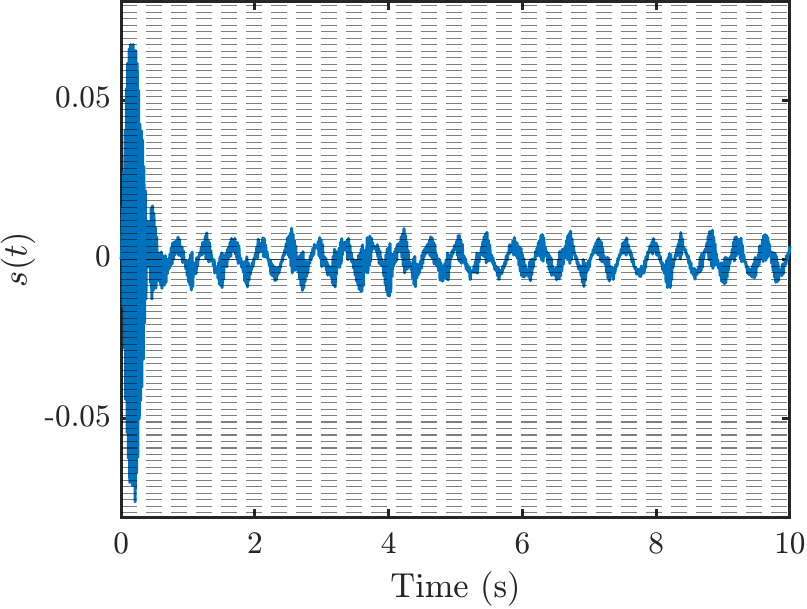}
\caption{Performance under reduced amplitude perturbation for $T_s = 1e^{-2}$ and $N = 50$.}
\label{fig:S_N_Ts_recovery_scenario}
\end{figure}

Furthermore, it is demonstrated that even for a relatively large sampling time ($T_s = 1e^{-2}$), satisfactory performance can be recovered by selecting an appropriate number of barriers. As shown in Fig.~\ref{fig:S_N_Ts_recovery_scenario}, the choice of $N = 50$ ensures that the system response remains within the maximum acceptable error bounds. This result highlights the robustness of the proposed approach and its capability to maintain safe operation despite adverse sampling conditions.

\subsection{Discussion}
Overall, the simulation results demonstrate that the proposed discrete-time multi-layer barrier adaptive super-twisting controller achieves robust tracking performance in the presence of severe disturbances and sampling constraints. The conducted sensitivity analysis highlights the importance of appropriate parameter selection and confirms the effectiveness of combining barrier-based gain adaptation with eigenvalue-preserving discretization to recover continuous-time performance characteristics in digital control implementations.


\section{Conclusions}\label{sec:Conclusions}
In this paper, a discrete-time implementation of a multi-layer barrier function–based adaptive super-twisting controller has been presented. The proposed approach addresses the limitations arising from inter-sample blindness in digital sliding mode control by employing a nested barrier architecture that modulates the control gains across multiple regions of the sliding variable. A discretization strategy based on eigenvalue exact matching has been adopted to preserve the stability and robustness properties established in continuous time. It was shown that the resulting discrete-time controller retains the adaptive characteristics of the continuous-time design while ensuring boundedness of the closed-loop trajectories under unknown but bounded perturbations.



\addtolength{\textheight}{-12cm}   





\section*{ACKNOWLEDGMENTS}
This research was conducted at DTU Electro as part of the HyRel project, funded by the Danish Energy Technology Development and Demonstration Programm (EUDP), grant number 64022-1058. The authors appreciate their support. Leonid Fridman acknowledges the financial support of  the Programa de Apoyo a Proyectos de Investigacion e Innovacion Tecnologica, Direccion General de Asuntos del Personal Academico, Universidad Nacional Autonoma de Mexico under Project IT100625.


\balance
\bibliographystyle{IEEEtran}        
\bibliography{Bibliography/mybib}

\end{document}